\documentclass[12pt]{article}
\usepackage{amsfonts}
\usepackage{amsmath,amssymb,color}
\usepackage{graphicx}
\usepackage{subcaption}
\usepackage{array,multirow}
\usepackage{float}
\usepackage{afterpage}
\usepackage{makecell}

\usepackage{xcolor}
\usepackage{amsfonts}
\usepackage{amsmath,amssymb}
\usepackage{color}
\usepackage{graphicx}
\usepackage{subcaption}
\usepackage{array,multirow}
\usepackage{float}
\usepackage{afterpage}
\usepackage{arydshln}
\usepackage{mathtools}
\usepackage{natbib}
\usepackage[doublespacing]{setspace}
\usepackage{makecell}

\newcommand*{\QEDB}{\hfill\ensuremath{\square}}%

\newcommand*{\Cdot}{\raisebox{-0.25ex}{\scalebox{1.5}{$\cdot$}}}
\newcommand{\aaa}{\boldsymbol \alpha}

\newcommand{\ttt}{\boldsymbol \theta}

\newcommand{\rr}{\boldsymbol r}

\newcommand{\mo}{\mathbf 1}
\newcommand{\mz}{\mathbf 0}
\newcommand{\bq}{\boldsymbol q}
\newcommand{\cs}{\boldsymbol s}
\newcommand{\ca}{\boldsymbol x}
\newcommand{\cb}{\boldsymbol y}
\newcommand{\ba}{\boldsymbol a}
\newcommand{\bb}{\boldsymbol b}

\newcommand{\vv}{\boldsymbol v}

\newcommand{\ee}{\boldsymbol e}

\newcommand{\pp}{{\boldsymbol p}}

\newcommand{\qq}{\boldsymbol q}
\newcommand{\cg}{\mbox{$\boldsymbol g$}}

\newcommand{\RR}{\boldsymbol R}

\newtheorem{theorem}{Theorem}

\newtheorem{condition}{Condition}

\newtheorem{corollary}{Corollary}

\newtheorem{definition}{Definition}
\newtheorem{example}{Example}

\newtheorem{lemma}{Lemma}

\newtheorem{proposition}{Proposition}
\newtheorem{remark}{Remark}

\begin{document}
\onehalfspacing
\title{The Sufficient and Necessary Condition for the Identifiability and Estimability of the DINA Model}
\author{Yuqi Gu and Gongjun Xu \\  { University of Michigan}}
\date{}
\maketitle
\doublespacing

\begin{abstract}
	Cognitive Diagnosis Models (CDMs) are useful statistical tools in cognitive diagnosis assessment. However, as many other latent variable models, the CDMs often suffer from the non-identifiability issue. 
This work gives the sufficient and necessary condition for identifiability of the basic DINA model, which not only addresses the open problem in Xu and Zhang (2016, {\it Psychomatrika}, {\bf 81}:625-649) on the minimal requirement for identifiability, 
 but also sheds light on the study of more general CDMs, which often cover DINA as a submodel. 
Moreover, we show the identifiability condition  ensures the consistent estimation of the model parameters.
From a practical perspective, the identifiability condition only depends on the $Q$-matrix structure and is easy to verify, which would provide a guideline for designing statistically valid and estimable cognitive diagnosis tests.
\end{abstract}

\section{Introduction}

Cognitive Diagnosis Models (CDMs), also called    Diagnostic Classification Models (DCMs),  are useful statistical tools in cognitive diagnosis assessment, which aims to achieve a fine-grained decision on an individual's latent attributes, such as skills, knowledge, personality traits, or psychological disorders, based on his or her observed responses to some designed diagnostic items. The CDMs fall into the more general regime of restricted latent class models in the statistics literature, and model the complex relationships among the items, the  latent attributes and the  item responses for a set of items and a sample of respondents. Various CDMs have been developed with different cognitive diagnosis assumptions, among which the Deterministic Input Noisy output ``And" gate model \citep[DINA;][]{Junker} is a popular one and serves as a basic submodel for more general CDMs such as the general diagnostic model \citep{von}, the log linear CDM \citep[LCDM;][]{HensonTemplin09}, and the generalized DINA  model \citep[GDINA;][]{dela2011}.

To achieve reliable and valid diagnostic assessment, a fundamental issue is to ensure that the CDMs applied in the cognitive diagnosis are statistically identifiable, which is a necessity for   consistent estimation of the model parameters of interest and   valid statistical inferences.  
 The study of identifiability in statistics and psychometrics has a long history \citep{Koopmans50,McHugh,rothenberg1971identification,GOODMAN74,gabrielsen1978consistency}.
 The identifiability issue  of the CDMs has also long been a concern, as noted in the literature   \citep*{DiBello,MarisBechger,TatsuokaC09,deCarlo2011,davier2014dina}. 
 In practice, however, there is often a tendency to overlook the   issue due to the lack of easily verifiable identifiability conditions. Recently there have been several  studies on the identifiability of the CDMs, including the DINA model \citep[e.g.,][]{Xu15} and  general models \citep[e.g.,][]{Xu2016,Xu2017,GLY}.

  However, the existing works mostly focus  on developing  sufficient conditions for the 
model identifiability, which might impose stronger than needed or sometimes  even impractical constraints on designing identifiable cognitive diagnostic tests.
It remains an open problem in the literature what would be the minimal requirement, i.e., the sufficient and necessary conditions, for the models  to be identifiable. 
In particular, for the DINA model, \cite{Xu15} proposed a set of sufficient conditions and a set of necessary conditions for the identifiability of the slipping, guessing and population proportion  parameters. However, as pointed out by the authors, there is a gap between the two sets of conditions;  
 see \cite{Xu15} for examples and discussions.

This paper addresses this open problem by developing the sufficient and necessary condition for the identifiability of the DINA model.
Furthermore, we show that the identifiability condition ensures the statistical consistency of the maximum likelihood estimators of the   model parameters.
 The proposed condition not only  guarantees the identifiability, but also  gives the minimal requirement that the DINA model needs to meet in order to be identifiable. 
The identifiability result can be directly applied to the DINO model \citep{Templin} through the duality of the DINA and DINO models. 
For general CDMs such as the LCDM and GDINA models,  since the DINA model can be considered as a submodel of them, the proposed   condition    also serves as a necessary requirement.  
From a practical perspective, the sufficient and necessary condition only depends on the $Q$-matrix structure and such  easily checkable condition  would provide a practical guideline for designing statistically valid and estimable cognitive tests.

The rest of the paper is organized as follows. Section 2 introduces the basic model setup and the definition of identifiability. Section 3 states the identifiability results and includes several illustrating examples. Section 4 gives a further discussion and the Appendix provides the proof  of the main results.

\section{Model Setup and Identifiability}

We consider the setting of a cognitive diagnosis test with binary responses. The test contains $J$ items to measure $K$ unobserved latent attributes. 
The latent attributes are assumed to be binary for diagnosis purpose and 
a complete configuration of the $K$ latent attributes is called   an attribute profile, which is denoted by a $K$-dimensional binary vector $\aaa=(\alpha_1,\ldots,\alpha_K)^\top$, where $\alpha_k\in\{0,1\}$ represents deficiency or mastery of the $k$th   attribute.
The underlying cognitive structure, i.e. the relationship between the items and the attributes, is described by the so-called $Q$-matrix, originally proposed by \cite{Tatsuoka1983}. A $Q$-matrix $Q$ is a $J\times K$ binary matrix with entries $q_{j,k}\in\{0,1\}$ indicating the absence or presence of the dependence of the $j$th item on the $k$th attribute. The $j$th row vector $\qq_j$ of the $Q$-matrix, also called the $\qq$-vector, corresponds to the attribute requirements of item $j$.  

A subject's attribute profile is assumed to  follow a categorical  distribution with population proportion parameters $\pp := (p_{\aaa}:\aaa\in\{0,1\}^K)^\top$,
where $p_{\aaa}$ is the proportion of attribute profile $\aaa$ in the population and $\pp$ satisfies  $\sum_{\aaa\in\{0,1\}^K}p_{\aaa}=1$ and $p_{\aaa}>0$ for any $\aaa\in\{0,1\}^K$. 
For an attribute profile $\aaa$ and a $\qq$-vector $\qq_j$, 
  we write $\aaa\succeq \qq_j$   if $\aaa$ masters all the required attributes of item $j$, i.e., $\alpha_k \geq q_{j,k}$ for any $k \in\{1, \ldots, K\}$, and  write $\aaa\nsucceq \qq_j$ if there exists some $k$ such that $\alpha_k < q_{j,k}.$ Similarly we define the operations $\preceq$ and $\npreceq$.

A subject provides a $J$-dimensional binary response vector $\RR =(R_1,\ldots,R_J)^\top\in\{0,1\}^J$ to these $J$ items. 
The DINA model assumes a conjunctive relationship among attributes, which means it is necessary to master all the attributes required by an item to be capable of providing a positive response to it. Moreover, mastering additional unnecessary attributes does not compensate for the lack of the necessary attributes. 
 Specifically,  for any item $j$ and attribute profile $\aaa$, we define  the binary ideal response $\xi_{j,\aaa}  = I(\aaa\succeq\qq_j)$.
If there is no uncertainty in the response, then a subject with attribute profile $\aaa$ will  have response $R_j=\xi_{j,\aaa}  $ to item $j$.
The uncertainty of the responses is incorporated at the item level, using slipping and guessing parameters. 
For each item $j$, the slipping parameter 
$s_j := P(R_j=0\mid \xi_{j,\aaa} =1)$ denotes the probability of a subject giving a negative response despite mastering all the necessary skills; while the guessing parameter
$g_j := P(R_j=1\mid \xi_{j,\aaa} =0)$
denotes the probability of giving a positive response despite deficiency of some necessary skills. 

Note that if  some item $j$ does not require any of the attributes, namely $\bq_j$ equals the  zero vector $\mathbf 0$, then $\xi_{j,\aaa}=1$ for all attribute profiles $\aaa\in\{0,1\}^K$. Therefore, in this special case, the guessing parameter   is not needed in the specification of the DINA model. 
The DINA model item parameters then include  slipping parameters $\cs = (s_1,\ldots,s_J)^\top$ 
and guessing parameters 
$\cg^{-} = (g_j: \forall j \mbox{ such that } \bq_j\neq  \mathbf 0 )^\top$. 
We assume $1-s_j>g_j$ for any item $j$ with $\bq_j\neq  \mathbf 0$.
For notational simplicity, in the following discussion we define  the guessing parameter of any item with $\bq_j=\mathbf 0$ to be a   known value  $g_j\equiv 0$, and  write   $\cg=(g_1,\ldots, g_J)^\top$.

Conditional on the attribute profile $\aaa$, the DINA model further assumes a subject's responses are   independent. Therefore the probability mass function of a subject's response vector $\RR = (R_1,\ldots,R_J)^\top$ is
 \begin{equation}\label{prob}
 P(\RR=\rr\mid Q,\cs,\cg,\pp) 
 =  \sum_{\aaa\in\{0,1\}^K}p_{\aaa} \prod_{j=1}^J (1-s_j)^{\xi_{j,\aaa} r_j} g_j^{(1-\xi_{j,\aaa})r_j} s_j^{\xi_{j,\aaa}(1-r_j)}(1-g_j)^{(1-\xi_{j,\aaa})(1-r_j)},
\end{equation} 
where $\rr= (r_1,\ldots,r_J)^\top\in \{0,1\}^J$.

Suppose we have $N$ independent subjects, indexed by $i=1,\ldots, N$, in a cognitive diagnostic assessment.
We denote their response vectors by $\{\RR_i: i=1,\ldots,N\}$, which are our observed data.  
 The DINA model parameters that we aim to estimate from the  response data are $(\cs,\cg,\pp)$, based on which we can further evaluate the subjects' attribute profiles from their ``posterior distributions". 
To consistently estimate $(\cs,\cg,\pp)$, we need them to be identifiable.
Following the definition in the statistics literature \citep[e.g.,][]{Casella}, we say a set of parameters in the parameter space $B$ for a family of  probability density (mass)  functions $\{f(~\cdot\mid\beta):\beta\in B\}$ is identifiable if distinct values of $\beta$ correspond to distinct $f(~\cdot \mid\beta)$ functions, i.e., for any $\beta$ there is no $\tilde\beta\in B\backslash\{\beta\}$ such that $f(~\cdot\mid\beta) \equiv f(~\cdot\mid\tilde\beta)$.
In the context of the DINA model, we have the following definition. 
\begin{definition}
We say the DINA model parameters   are identifiable if there  is no $(\bar\cs,\bar\cg,\bar\pp)\neq (\cs,\cg,\pp)$ such that
\begin{equation}\label{eq-orig}
P(\RR=\rr\mid Q,\cs,\cg,\pp) = P(\RR=\rr\mid Q,\bar\cs,\bar\cg,\bar\pp) \mbox{ for all }  \rr\in \{0,1\}^J. 
\end{equation}
\end{definition}

\begin{remark}\label{rmk-idf}
Identifiability of latent class models is a well established concept in the literature \citep[e.g.,][]{McHugh,GOODMAN74}.
Recent studies on the identifiability of the CDMs and the restricted latent class models include \cite{JLGXZY2011}, \cite{Chen2014}, \cite{Xu15}, \cite{Xu2016}, and \cite{Xu2017}. 
However, as discussed in the introduction, most of them focus on developing   sufficient conditions while the sufficient and necessary conditions are still unknown. 
\end{remark}

\section{Main Result }
 We first introduce the important concept of the completeness of a $Q$-matrix, which was first introduced in \cite{Chiu}.
 A $Q$-matrix is said to be complete if  it can differentiate all latent attribute profiles, in the sense that under the $Q$-matrix, different attribute profiles have different response distributions. In this study of the DINA model, completeness of the $Q$-matrix means that $\{\ee_k^\top:k=1,\ldots,K\}\subseteq \{\qq_j:j=1,\ldots,J\}$, equivalently, for each attribute there is some item which requires that and solely requires that attribute.  Up to some row permutation, a complete $Q$-matrix  under the DINA model contains a $K\times K$ identity matrix.
Under the DINA model, completeness of the $Q$-matrix is necessary for identifiability of the population proportion parameters $\pp$ \citep{Xu15}.

Besides the completeness, an additional necessary condition for identifiability was also specified in \cite{Xu15} that each attribute needs to be related with at least 3 items.
For easy discussion, we summarize the set of necessary conditions in  \cite{Xu15} as follows.
 \begin{condition}
 \begin{itemize} 
\item[(i)] The $Q$-matrix is complete {under the DINA model} and without loss of generality, we assume the $Q$-matrix takes the following form:
\begin{equation}\label{eq-formq}
Q=\left(\begin{array}{c}
\mathcal I_K \\
\hdashline[2pt/2pt]
Q^*
\end{array}\right)_{J\times K},
\end{equation}
where  ${\cal I}_K$ denotes the $K\times K$ identity matrix and $Q^*$ is a $(J-K)\times K$ submatrix of $Q$.

\item[(ii)] Each of the $K$ attributes is required by at least 3 items.
\end{itemize}
 \end{condition}
Though necessary, \cite{Xu15} recognized that Condition 1 is not sufficient. To establish identifiability, the authors  also  proposed a set of sufficient conditions, which however is not necessary. 
 For instance,   the  $Q$-matrix in \eqref{ex1}, which is given on   page 633 in \cite{Xu15}, does not satisfy their sufficient condition but still gives an identifiable model.

\singlespacing
\begin{equation}\label{ex1}
	Q =\begin{pmatrix}
&{\cal I}_4&   \\
\hdashline[2pt/2pt]
1&1\quad1&0\\
1&1\quad0&1\\
1&0\quad1&1\\
0&0\quad0&1
\end{pmatrix} 
\end{equation}

	\vspace{3mm}
\doublespacing
\noindent
In particular, their sufficient condition C4 requires that for each $k\in\{1,\ldots, K\}$,  there exist two subsets $S_k^+$ and $S_k^-$ of the items (not necessarily   nonempty or disjoint) in $Q^*$ such that $S_k^+$ and $S_k^-$ have attribute requirements that are identical except in the $k$th attribute, which is required by an item in $S_k^+$ but not by any item in $S_k^-$.  However, the first attribute in \eqref{ex1} does not satisfy this condition. 
Examples of this kind of $Q$-matrices not satisfying their C4 but still identifiable are not rare and can be easily constructed as shown below in \eqref{ex2}.

\singlespacing
\begin{equation}\label{ex2}
Q =\begin{pmatrix}
&{\cal I}_3&\\
\hdashline[2pt/2pt]
1&1&0 \\
1&0&1 \\
1&1&1 \\
1&1&1 \\
\end{pmatrix},\quad 
Q =\begin{pmatrix}
&{\cal I}_3&\\
\hdashline[2pt/2pt]
1&0&0 \\
1&1&0 \\
1&1&1 \\
0&0&1 \\
\end{pmatrix},\quad 
Q =\begin{pmatrix}
&{\cal I}_3&\\
\hdashline[2pt/2pt]
1&0&0 \\
1&1&0 \\
1&1&1 \\
1&1&1 \\
\end{pmatrix},\quad 
Q =\begin{pmatrix}
&{\cal I}_4&\\
\hdashline[2pt/2pt]
1&1\quad1&0\\
1&1\quad0&1\\
1&0\quad1&1\\
0&1\quad0&1
\end{pmatrix}.
	\end{equation}
	
	\vspace{3mm}
\doublespacing
	
It has been an open problem in the literature what would be the minimal requirement of the $Q$-matrix for the model to be identifiable. 
This paper solves this   problem and shows shat Condition 1 together with the following Condition 2 are sufficient and necessary for the identifiability of the DINA model parameters.
\begin{condition} Any two different columns of the sub-matrix $Q^*$ in \eqref{eq-formq} are distinct.
\end{condition}
We have the following identifiability result.
\begin{theorem}[Sufficient and Necessary Condition]
\label{thm-com-necc}
Conditions 1 and 2 are sufficient and necessary for the identifiability of all the DINA model parameters.
\end{theorem}

\begin{remark}\label{rmk-0}
 From the model construction, when there are some items that require none of the attributes, all the DINA model parameters are $(\cs,\pp)$ and $\cg^{-} = (g_j: \forall j \mbox{ such that } \bq_j\neq  \mathbf 0 )^\top$.
Theorem \ref{thm-com-necc} also applies to this special case   that the proposed conditions still remain sufficient and necessary for the identifiability of $(\cs,\cg^{-},\pp)$, under a  $Q$-matrix containing some all-zero $\bq$-vectors. See   Proposition \ref{prop-q0} in the Appendix  for more details.
\end{remark}

\color{black}
Conditions 1 and 2 are easy to verify.  Based on Theorem 1,  it is recommended in practice to design the $Q$-matrix such that it is complete,  has each attribute required by at least 3 items, and has $K$ distinct columns in the sub-matrix $Q^*$.
Otherwise, the model parameters would suffer from the non-identifiability issue.   
 We use the following examples to illustrate the theoretical result. 

\begin{example}
 From Theorem 1, the $Q$-matrices in \eqref{ex1} and \eqref{ex2} satisfy both Conditions 1 and 2  and therefore give identifiable models, while the results in  \cite{Xu15} cannot be applied since their condition C4 does not hold. 
On the other hand, the $Q$-matrices below in \eqref{ex3} satisfy the necessary conditions in \cite{Xu15}, but they do not satisfy our Condition 2, so the corresponding models are not identifiable.

\singlespacing
\begin{equation}\label{ex3}
Q =\begin{pmatrix}
&{\cal I}_3&\\
\hdashline[2pt/2pt]
1&1&1 \\
1&1&1 \\
1&1&1 \\
1&1&1 \\
\end{pmatrix},\quad 
Q =\begin{pmatrix}
&{\cal I}_3&\\
\hdashline[2pt/2pt]
1&1&0 \\
1&1&0 \\
0&0&1 \\
0&0&1 \\
\end{pmatrix},\quad 
Q =\begin{pmatrix}
&{\cal I}_3&\\
\hdashline[2pt/2pt]
1&1&0 \\
1&1&1 \\
0&0&1 \\
0&0&1 \\
\end{pmatrix},\quad 
Q =\begin{pmatrix}
&{\cal I}_4&\\
\hdashline[2pt/2pt]
1&1\quad1&0\\
1&1\quad1&1\\
1&0\quad1&1\\
0&1\quad0&1
\end{pmatrix}.
	\end{equation}
\end{example}

\vspace{3mm}
\doublespacing

\begin{example}  
 To  illustrate the necessity of Condition 2, we consider a simple case when $K=2$. 
 If Condition 1 is satisfied but Condition 2 does not hold, the $Q$-matrix can only have the following form  up to some row permutations,

\vspace{-4mm}
\singlespacing
\begin{equation}\label{exC2}
Q =\begin{pmatrix}
 {\cal I}_2  \\
 \hdashline[2pt/2pt]
0\quad 0\\
\vdots\quad \vdots\\
0\quad 0\\
1\quad 1\\
\vdots\quad \vdots\\
1\quad 1\\
\end{pmatrix}_{J\times 2},
 \end{equation}
 
\doublespacing
\noindent 
 where the first two items give an identity matrix while the next $J_0$ items require none of the attributes and the last $J-2-J_0$ items require both attributes. 
Under the $Q$-matrix in \eqref{exC2}, we next show the model parameters $(\cs,\cg,\pp)$ are not identifiable by constructing a set of parameters   $(\bar\cs,\bar\cg,\bar\pp)\neq (\cs,\cg,\pp)$ which satisfy \eqref{eq-orig}. 
Recall from the model setup in Section 2 that for any item   $j\in\{3,\ldots,J_0+2\}$ that has   $\bq_j=\mathbf 0$, the guessing parameter   is not needed by the DINA model and for notational convenience, we set $g_j \equiv\bar g_j \equiv 0$.  
We take  
$\bar \cs =\cs$, $\bar g_j=g_j$ for $j=J_0+3,\ldots, J$, and $\bar p_{(11)}=p_{(11)}$. Next we show the remaining parameters $(g_1,g_2,p_{(00)}, p_{(10)},p_{(01)})$ are not identifiable. 
From  Definition 1,   the non-identifiability occurs if the following equations hold (see the Supplementary Material for the computational details):
$P\big((R_1,R_2)=(r_1,r_2)\mid Q, \bar\cs,\bar\cg,\bar\pp\big) =  P\big((R_1,R_2)=(r_1,r_2)\mid Q, \cs,\cg,\pp\big)$ for all $(r_1,r_2)\in\{0,1\}^2,$
where $(R_1,R_2)$ are the first two entries of the random response vector $\RR$.
These equations can be further expressed as the following equations in \eqref{a^2}:

\vspace{-4mm}
\onehalfspacing
\begin{equation}\label{a^2}
(r_1,r_2) =
\begin{cases}
(0,0):~ \bar p_{(00)} + \bar p_{(10)} + \bar p_{(01)} + p_{(11)}
	= p_{(00)} + p_{(10)} + p_{(01)}+ p_{(11)};\\
(1,0):~\bar g_1[\bar p_{(00)} +  \bar p_{(01)}] + (1-s_1)[\bar p_{(10)} +   p_{(11)}] \\
	\qquad\qquad
	= g_1[ p_{(00)} +   p_{(01)}] + (1-s_1)[ p_{(10)} +   p_{(11)}];\\
(0,1):~\bar g_2[\bar p_{(00)} +  \bar p_{(10)}] + (1-s_2)[\bar p_{(01)} +   p_{(11)}] \\
	\qquad\qquad
	= g_2[ p_{(00)} +   p_{(10)}] + (1-s_2)[ p_{(01)} +   p_{(11)}];\\
(1,1):~		\bar g_1\bar g_2\bar p_{(00)} + \bar g_1(1- s_2)\bar p_{(01)}+(1- s_2)\bar g_2\bar p_{(10)}+(1- s_1)(1- s_2) p_{(11)} \\
	\qquad\qquad	= g_1 g_2\bar p_{(00)} +  g_1(1- s_2) p_{(01)}+(1- s_2) g_2 p_{(10)}+(1- s_1)(1- s_2) p_{(11)} .
\end{cases}
\end{equation}
\doublespacing
\color{black}
For any  $(\cs,\cg,\pp)$, there are 4 constraints in \eqref{a^2}
but   5 parameters 	$(\bar g_1,\bar g_2,\bar p_{(00)}, \bar p_{(10)},\bar p_{(01)})$ to solve. Therefore there are infinitely many solutions and $(\cs,\cg,\pp)$ are non-identifiable.
\end{example}  

\begin{example}  We provide a numerical illustration  of Example 2. 
Without loss of generality, we take $J_0=0$,   since whether there exist zero $\bq$-vector items makes no impact on the nonidentifiability phenomenon as illustrated in \eqref{a^2}.
We take $J=10$ and set the true parameters to be $(p_{(00)},~p_{(10)},~p_{(01)},~p_{(11)})=(0.1,~ 0.3,~ 0.4,~ 0.2)$ and $s_j=g_j=0.2$ for $j\in\{1,\ldots, 10\}$. We first generate a random sample of size $N=200$. From the data, we obtain one set of maximum likelihood estimators  as follows:
\[\begin{aligned}
&(\hat p_{(00)},~\hat p_{(10)},~\hat p_{(01)},~\hat p_{(11)}) =(\mathbf{0.22346},~\mathbf{0.26298},~\mathbf{0.32847},~0.18509);\\
&\hat\cs = (0.1269,~0.1541,~0.0000,~0.2015,~0.1549,~0.2638,~0.3551,~0.1903,~0.1843,~0.1468);\\
&\hat\cg = (\mathbf{0.1678},~\mathbf{0.2011},~ 0.2330,~ 0.1990,~ 0.2007,~ 0.2316,~ 0.2155,~ 0.1720,~ 0.2197,~ 0.1805).
\end{aligned}\]
Based on \eqref{a^2}, we can construct infinitely many sets  of  $(\bar\cs,\bar\cg,\bar\pp)$ that are also maximum likelihood estimators.   For instance, we take $\bar\cs=\hat\cs$, $\bar g_j = \hat g_j$ for $j=3,\ldots,10$, $\bar p_{(11)}=\hat p_{(11)}$, and $\bar p_{(00)} = 0.998\cdot \hat p_{(00)}$. Then solve  \eqref{a^2}  for the remaining parameters $\bar p_{(10)}$, $\bar p_{(01)}$, $\bar g_1$ and $\bar g_2$ to get
\[
\bar p_{(00)} = \mathbf{0.22301},\quad
\bar p_{(01)} = \mathbf{0.33306},\quad
\bar p_{(10)} = \mathbf{0.25884},\quad
\bar g_1 = \mathbf{0.2561},\quad
\bar g_2 = \mathbf{0.1073}.
\]
The two different sets of values $(\hat\cs,\hat\cg,\hat\pp)$ and $(\bar\cs,\bar\cg,\bar\pp)$ both give the identical log-likelihood value -1132.1264, which confirms the non-identifiablility.

To further illustrate the above argument does not depend on the sample size, we  generate a random sample of size $N=10^5$    and obtain the following  estimators:  
\[\begin{aligned}
&(\hat p_{(00)},~\hat p_{(10)},~\hat p_{(01)},~\hat p_{(11)}) =(\mathbf{0.10436},~\mathbf{0.29933},~\mathbf{0.39845},~0.19786);\\
&\hat\cs = (0.1968  ,~  0.1932 ,~   0.2007 ,~   0.2065 ,~   0.2015  ,~  0.2000  ,~  0.2001 ,~   0.1949  ,~  0.1985 ,~   0.2036);\\
&\hat\cg = (\mathbf{0.1993},~ \mathbf{0.2006},~ 0.1995  ,~   0.2010  ,~   0.1971   ,~  0.1983  ,~   0.1995 ,~    0.2022 ,~    0.1989 ,~    0.1988).
\end{aligned}\]     
Similarly, we set   $\bar\cs=\hat\cs$, $\bar g_j = \hat g_j$ for $j=3,\ldots,10$, $\bar p_{(11)}=\hat p_{(11)}$, and $\bar p_{(00)} = 0.998\cdot \hat p_{(00)}$. Solving \eqref{a^2}   gives
\[
\bar p_{(00)} = \mathbf{0.10415},\quad
\bar p_{(01)} = \mathbf{0.40161},\quad
\bar p_{(10)} = \mathbf{0.29638},\quad
\bar g_1 = \mathbf{0.3212},\quad
\bar g_2 = \mathbf{0.0458}.
\]
where the two different sets of values $(\hat\cs,\hat\cg,\hat\pp)$ and $(\bar\cs,\bar\cg,\bar\pp)$ both lead to the identical log-likelihood value -571659.1708.
This illustrates that the non-identifiability issue depends on the model setting instead of the sample size. In practice, as long as Conditions 1 and 2 do not hold, we may suffer from similar non-identifiability issues no matter how large the sample size is. 
\end{example}  
 

Identifiability is the prerequisite and a necessary condition for  consistent estimation.   Here we say a parameter is consistently estimable if we can construct a consistent estimator for the parameter. That is, for parameter $\beta$, there exists $\hat\beta_N$ such that $\hat\beta_N-\beta\to 0$ in probability as the sample size $N\to\infty$.
When the identifiability conditions are satisfied, we show that  the maximum likelihood estimators (MLEs) of the DINA model parameters $(\cs,\cg,\pp)$ are statistically consistent as   $N\to\infty$.
For the observed responses $\{\RR_i: i=1,\ldots, N\}$, we can write their   likelihood function as
\begin{equation}\label{eq-ll}
L_N(\cs,\cg,\pp;\,\RR_1,\ldots,\RR_N) = \prod_{i=1}^N P(\RR=\RR_i\mid Q,\cs,\cg,\pp),
\end{equation}
where $P(\RR=\RR_i\mid Q,\cs,\cg,\pp)$ is as defined in \eqref{prob}. Let $(\hat\cs, \hat\cg,\hat\pp)$ be the corresponding MLEs  based on   (\ref{eq-ll}).
We have the following corollary.
\begin{corollary}\label{cor1}
When Conditions 1 and 2 are satisfied, the MLEs    $(\hat\cs, \hat\cg,\hat\pp)$ are consistent as $N\to\infty$.
\end{corollary}

The results in Theorem 1 and Corollary 1 can be directly applied to the DINO model through the duality of the DINA and DINO models \citep[see Proposition 1 in][]{Chen2014}.
Specifically, when Conditions 1 and 2 are satisfied, the guessing, slipping, and population proportion parameters in the DINO model are identifiable and can also be consistently estimated as $N\to\infty$.

Moreover, the proof of Corollary \ref{cor1} can be directly generalized to the other CDMs that the MLEs of the model parameters, including the item parameters and population proportion parameters, are consistent as $N\to\infty$ if they are identifiable. 
Therefore under the sufficient conditions for identifiability of general CDMs developed in the literature such as \cite{Xu2016}, the model parameters are also consistently estimable.
 Although the minimal requirement for identifiability and estimability of general CDMs are still unknown, the proposed Conditions 1 and 2 are necessary since the DINA model is a submodel of them.
 For instance, \cite{Xu2016} requires two identity matrices in the $Q$-matrix to obtain identifiability, which automatically satisfies Conditions 1 and 2 in this paper.

 We  next present an example to illustrate that when the proposed conditions are satisfied, the MLEs of the DINA model parameters are consistent.
\begin{example}\label{exp-consist} 
We perform a simulation study with the following $Q$-matrix that satisfies the proposed sufficient and necessary conditions. The true parameters are set to be $p_{\aaa}=0.125$ for all $\aaa\in\{0,1\}^3$, and $s_j=g_j=0.2$ for  $j=1,\ldots,6$.

\vspace{-4mm}
\singlespacing
\begin{equation*}
Q=\begin{pmatrix}
1 & 0 & 0 \\
0 & 1 & 0 \\
0 & 0 & 1 \\
0 & 1 & 1 \\
1 & 0 & 1 \\
1 & 1 & 0 \\
\end{pmatrix},
\end{equation*}

\vspace{2mm}
\doublespacing
\noindent
For each sample size $N=200\cdot i$ where $i=1,\ldots,10$, we generate 1000 independent datasets, and use the EM algorithm with random initializations to obtain the MLEs of model parameters  for each dataset. The mean squared errors (MSEs) of the parameters $\cs$, $\cg$, $\pp$  computed from the 1000 runs are shown in Table \ref{tab-mse} and Figure \ref{fig-mse}. One can see that the MSEs keep decreasing as the sample size $N$ increases, matching the theoretical result in Corollary \ref{cor1}.
 
\begin{table}[h!]
\begin{center}
\begin{tabular}{|c|c|c|c|c|c|}
\hline
$N$ & 400 & 800 & 1200 & 1600 & 2000 \\
\hline
$\pp$ & 0.0272 & 0.0137 & 0.0087 & 0.0065 & 0.0051\\
$\cs$  & 0.0613 & 0.0335 & 0.0221 & 0.0174 & 0.0131\\
 $\cg$  &  0.0411  & 0.0224 & 0.0149 & 0.0109 & 0.0082\\
 \hline
 \end{tabular}
 \end{center}
\caption[]{MSEs of DINA Model Parameters}
\label{tab-mse}
\end{table}

\begin{figure}[H]
\begin{subfigure}{0.33\textwidth}
\includegraphics[width=\linewidth]{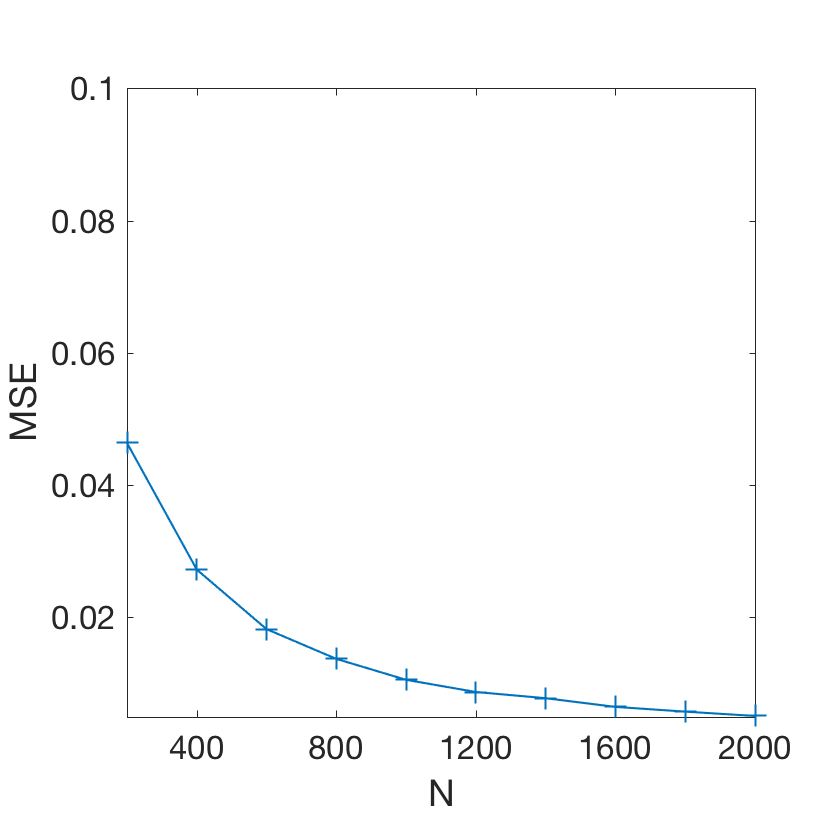}
\caption{MSEs of $\pp$} \label{fig:a}
\end{subfigure}\hspace*{\fill}
\begin{subfigure}{0.33\textwidth}
\includegraphics[width=\linewidth]{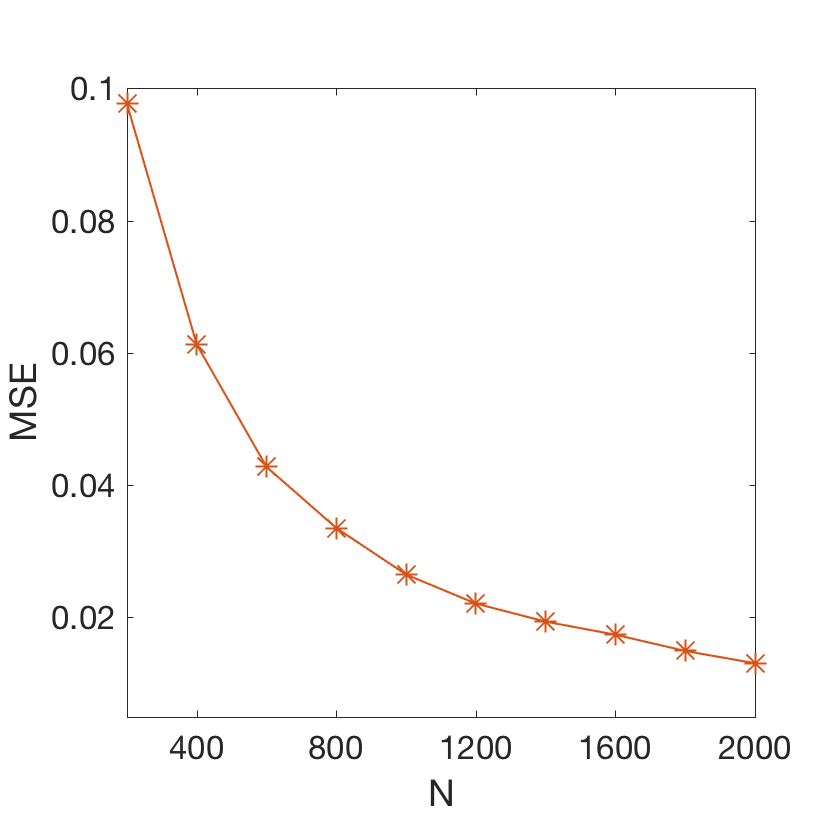}
\caption{MSEs of $\cs$} \label{fig:b}
\end{subfigure}\hspace*{\fill}
\begin{subfigure}{0.33\textwidth}
\includegraphics[width=\linewidth]{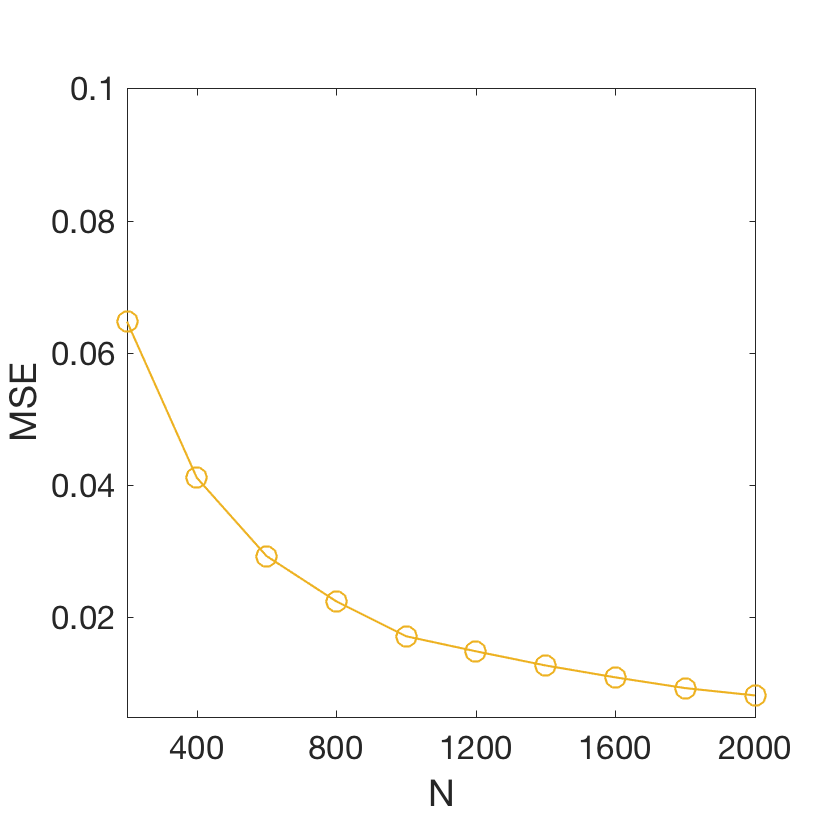}
\caption{MSEs of $\cg$} \label{fig:b}
\end{subfigure}
\caption[]{MSE of DINA Model Parameters versus Sample Size $N$}
\label{fig-mse}
\end{figure}

\end{example}
\color{black}

 \section{Discussion}
This paper presents the sufficient and necessary condition for identifiability of the DINA and DINO model parameters and establishes the consistency of the maximum likelihood estimators. As discussed in Section 3, the results would also shed light on the study of the sufficient and necessary conditions for general CDMs.

 This paper treats the attribute profiles as random effects from a population distribution. Under this setting, the identifiability conditions ensure the consistent estimation of the model parameters.
 However, generally in statistics and psychometrics, identifiability conditions are  not always sufficient for consistent estimation. 
 An example of identifiable but not consistently estimable is the fixed effects CDMs, where the subjects' attribute profiles are taken as model parameters. 
 Consider a simple example of the DINA model with nonzero but known slipping and guessing parameters. 
 Under the fixed effects setting, the model parameters include $\{\aaa_i, i=1,\ldots,N\}$, which are identifiable if the $Q$-matrix is complete \citep[e.g.,][]{Chiu}.
 But with fixed number of items, even when the sample size $N$ goes to infinity, the parameters $\{\aaa_i, i=1,\ldots,N\}$ cannot be consistently estimated.
 In this case, to have the consistent estimation of each $\aaa$, the number of items  needs to go to infinity and the number of identity sub-$Q$-matrices also needs to go to infinity \citep{wang2015consistency}, equivalently, there are infinitely many sub-$Q$-matrices satisfying Conditions 1 and 2. 

  When the identifiability conditions are not satisfied,   we may expect to obtain partial identification results that certain parameters are identifiable while others are only identifiable up to some transformations. For instance, when Condition 1 is satisfied, the slipping parameters are all identifiable and guessing parameters of items $(K+1,\ldots, J)$ are also identifiable. 
         It is also possible in practice that there  exist certain hierarchical structures among the latent attributes. For instance, an attribute may be a prerequisite for some other attributes. In this case, some entries of $\pp$ are restricted to be $0$. 
 It would also be interesting to consider the identifiability conditions under these restricted models.
For these cases, weaker conditions are expected for  identifiability of the model parameters. In particular, completeness of the $Q$-matrix may not be needed. We believe the techniques used in the proof of the main result can be extended to study such restricted models  and   would like to pursue this in the future.

\section*{Appendix: Proof of Theorem 1}
To study model identifiability, directly working with (\ref{eq-orig}) is technically challenging.
To facilitate the proof of the theorem, we introduce a key technical quantity following that of \cite{Xu2016}, the marginal probability matrix called the $T$-matrix. 
 The $T$-matrix $T(\cs,\cg)$, is a defined as a $2^J\times 2^K$ matrix, where the entries are indexed by row index $\rr\in \{0,1\}^J$ and column index $\aaa$.
Suppose that the columns of $T(\cs,\cg)$ indexed by $(\aaa^1,\ldots,\aaa^{2^K})$ are arranged in the following order of $\{0,1\}^K$ \[
\aaa^1 = \mz,~ \aaa^2=\ee_1,~ \ldots,~\aaa^{K+1} = \ee_K,~\aaa^{K+2} = \ee_1+\ee_2,~ \aaa^{K+3} = \ee_1+\ee_3,~ \ldots, ~\aaa^{2^K} = \sum_{k=1}^K\ee_k =\mo,
\]
where $\mz$ denotes the column vector of zeros, $\mo$ denotes the   column vector of  ones, and $\ee_k$ denotes a standard basis vector, whose $k$th element is one and the rest are zero; to simplify notation, we omit the dimension indices of $\mz, \mo$ and $\ee_k$'s.  Similarly, suppose that the  rows of $T(\cs,\cg)$ indexed by $(\rr^1,\ldots,\rr^{2^J})$ are arranged   in the following order
$$\rr^1 = \mz, ~ \rr^2=\ee_1,~ \ldots,~\rr^{J+1} = \ee_J,~\rr^{J+2} = \ee_1+\ee_2,~ \rr^{J+3} = \ee_1+\ee_3,~ \ldots, \rr^{2^J} = \sum_{j=1}^J\ee_j=\mo.$$
The $\rr=(r_1,\ldots, r_J)$th row and $\aaa$th column element of $T(\cs,\cg)$, denoted by $t_{\rr,\aaa}(\cs,\cg)$,  is  
the   probability that a subject with attribute profile $\aaa$ answers all items in the subset $\{j: r_j=1\}$ positively, that is,
 $t_{\rr,\aaa}(\cs,\cg) = P(\RR\succeq\rr\mid Q,\cs,\cg,\aaa). $
When $\rr=\mathbf 0$, 
$t_{\mz,\aaa}(\cs,\cg) = P(\rr\succeq \mz) = 1 \mbox{ for any } \aaa.$
When $\rr=\ee_j$, for $1\leq j\leq J$,  $t_{\ee_j,\aaa}(\cs,\cg) =P(R_j=1 \mid Q,\cs,\cg,\aaa).$ 
Let $T_{\rr,\Cdot}( \cs,\cg)$ be the row vector in the $T$-matrix corresponding to $\rr$.
Then for any $\rr \neq \mz$, we can write
$T_{\rr,\Cdot}( \cs,\cg) = \bigodot_{j: r_j=1} T_{\ee_j,\Cdot}( \cs,\cg),$ where $\odot$ is the element-wise product of the row vectors. 

 By definition, multiplying the $T$-matrix by the distribution of attribute profiles $\pp$ results in a vector, $T(\cs,\cg)\pp $, containing the marginal probabilities of successfully responding to each subset of items positively. The $\rr$th entry of this vector is 
\begin{eqnarray*}
T_{\rr,\Cdot}(\cs,\cg)\pp = \sum_{\aaa\in\{0,1\}^K} t_{\rr,\aaa}(\cs,\cg) p_{\aaa} 
=\sum_{\aaa\in\{0,1\}^K}  P(\RR\succeq \rr \mid Q,\cs,\cg,\aaa)p_{\aaa}= P(\RR\succeq \rr\mid Q,\cs,\cg,\pp).
\end{eqnarray*}
We can see that there is a one-to-one mapping between the two $2^J$-dimensional vectors $T(\cs,\cg)\pp$ and $\left(P(\RR= \rr\mid Q,\cs,\cg,\pp):~ \rr\in\{0,1\}^J\right)$. 
Therefore, Definition 1 directly implies the following proposition.  
\begin{proposition}\label{equiv}
The parameters $(\cs,\cg,\pp)$ are identifiable if and only if 
for any  $(\bar\cs,\bar\cg,\bar\pp)\neq (\cs,\cg,\pp)$, there exists $\rr\in \{ 0, 1\}^J$ 
such that 
\begin{equation}\label{gxx}
T_{\rr,\Cdot}( \cs,\cg) \pp\neq T_{\rr,\Cdot}(  \bar\cs,\bar\cg)\bar\pp.
\end{equation}
\end{proposition}
Proposition 1 shows that to establish the identifiability of  $(\cs,\cg,\pp)$, we only need to focus on the $T$-matrix structure.

 The following proposition characterizes the equivalence between  the identifiability of the DINA model associated with a $Q$-matrix with some zero $\bq$-vectors and that associated with the submatrix of $Q$ containing all of those nonzero $\bq$-vectors. The proof of Proposition \ref{prop-q0} is given in the Supplementary Material.
\begin{proposition}\label{prop-q0}
Suppose the $Q$-matrix of size $J\times K$ takes the form
\[Q=\begin{pmatrix}
Q'\\
\mathbf0\\
\end{pmatrix},\]
where $Q'$ denotes a $J'\times K$ submatrix containing the $J'$ nonzero $\bq$-vectors of $Q$, and $\mathbf0$ denotes a $(J-J')\times K$ submatrix containing those zero $\bq$-vectors of $Q$. Then the DINA model associated with $Q$ is identifiable if and only if the DINA model associated with $Q'$ is identifiable.
\end{proposition}
By Proposition \ref{prop-q0}, without loss of generality, in the following  we assume the $Q$-matrix   does not contain any zero $\bq$-vectors and  prove the necessity and sufficiency of the proposed Conditions 1 and 2.

\color{black}
\paragraph{Proof of Necessity}
The necessity of Condition 1 comes from Theorem 3 in \cite{Xu15}. Now suppose Condition 1 holds but Condition 2 is not satisfied. Without loss of generality, suppose the first two columns in $Q^*$ are the same  and the $Q$ takes the following form
\begin{equation}\label{eq-q1}
Q = \begin{pmatrix}
 	   & {\cal I}_K &     \\
  \hdashline[2pt/2pt]
 \vv & \vv \quad \vdots \quad  \vdots & \vdots 
 \end{pmatrix}_{J\times K},
\end{equation}
where $\vv$ is any binary vector of length $J-K$.
To show the necessity of Condition 2, 
 from Proposition 1, we only need to  find two different sets of parameters $(\cs,\cg,\pp)\neq ( \bar \cs,\bar\cg,\bar\pp)$ such that 
for any    $\rr \in\{0,1\}^J$, the following equation holds
\begin{equation}\label{eq-tpr}
T_{\rr,\cdot}( \cs,\cg)\pp = T_{\rr,\cdot}( \bar\cs,\bar\cg)\bar\pp.
\end{equation} 
We next construct such $(\cs,\cg,\pp)$ and $(\bar\cs, \bar\cg,\bar\pp)$. 
We assume in the following that $\bar \cs=\cs$ and $\bar g_j =g_j$ for any $j> 2$, and focus on the construction of  $(\bar g_1,\bar g_2,\bar\pp)\neq (  g_1,  g_2, \pp) $   satisfying \eqref{eq-tpr} for any $\rr \in\{0,1\}^J$.
For notational convenience, we write the positive response probability for item $j$ and attribute profile $\aaa$ in the following general form
$
\theta_{j,\aaa} := (1-s_j)^{\xi_{j,\aaa} }g_j^{1-\xi_{j,\aaa}}.
$
So based on our construction, for any $j>2$, $\theta_{j,\aaa} = \bar\theta_{j,\aaa}$.

We define two subsets of items $S_0$ and $S_1$ to be
\[
S_0 = \{j: q_{j,1}=q_{j,2}=0\} \mbox{ and }
S_1 = \{j: q_{j,1}=q_{j,2}=1\},
\]
where $S_0$ includes those items not requiring any of the first two attributes, and $S_1$ includes those items requiring both of the first two attributes.
Then since Condition 2 is not satisfied, we must have $S_0\cup S_1 = \{3,4,\ldots,J\}$, i.e., all but the first two items either fall in $S_0$ or $S_1$. Now consider any $\aaa^*\in\{0,1\}^{K-2}$, for any item $j\in S_0$, the four attribute profiles $(0,0,\aaa^*)$, $(0,1,\aaa^*)$, $(1,0,\aaa^*)$ and $(1,1,\aaa^*)$ always have the same positive response probabilities to $j$, and for any $j\in S_1$, the three attribute profiles $(0,0,\aaa^*)$, $(1,0,\aaa^*)$, $(0,1,\aaa^*)$ always have the same positive response probabilities to $j$. In summary, 
 \begin{eqnarray}\label{thetaeq}
 \begin{cases}
   \theta_{j,\,(0,0,\aaa^*)}  = \theta_{j,\,(0,1,\aaa^*)} = \theta_{j,\,(1,0,\aaa^*)} =\theta_{j,\,(1,1,\aaa^*)}& \mbox{ for }  j\in S_0;\\
 \theta_{j,\,(0,0,\aaa^*)}  = \theta_{j,\,(0,1,\aaa^*)} = \theta_{j,\,(1,0,\aaa^*)} \leq \theta_{j,\,(1,1,\aaa^*)}& \mbox{ for } j\in S_1.
 \end{cases}
 \end{eqnarray}

For any response vector $\rr\in\{0,1\}^J$ such that $\rr_{S_1}:=(r_j: j\in S_1)\neq\mz$, namely $r_j=1$ for some item $j$ requiring both of the first two attributes,
we discuss the following four cases.
\begin{enumerate}
\item[(a)]  For any $\rr$ such that $(r_1,r_2) = (0,0)$ and $\rr_{S_1}\neq\mz$, from \eqref{thetaeq} and the definition of the $T$-matrix,
  (\ref{eq-tpr}) is equivalent to
\[\begin{aligned}
&\sum_{\aaa^*}\biggr\{ \Big[\prod_{j>2:\,r_j=1} \theta_{j,\,(0,0,\aaa^*)}\Big] \big[p_{(0,0,\aaa^*)}+p_{(0,1,\aaa^*)}+p_{(1,0,\aaa^*)}\big]
+ \Big[\prod_{j>2:\,r_j=1} \theta_{j,\,(1,1,\aaa^*)}\Big] p_{(1,1,\aaa^*)} \biggr\} \\
=~&
\sum_{\aaa^*}\biggr\{ \Big[\prod_{j>2:\,r_j=1}\bar \theta_{j,\,(0,0,\aaa^*)}\Big]\big[\bar p_{(0,0,\aaa^*)}+\bar p_{(0,1,\aaa^*)}+\bar p_{(1,0,\aaa^*)}\big]
+ \Big[\prod_{j>2:\,r_j=1}\bar \theta_{j,\,(1,1,\aaa^*)}\Big]\bar p_{(1,1,\aaa^*)} \biggr\} \\
=~&
\sum_{\aaa^*}\biggr\{ \Big[\prod_{j>2:\,r_j=1} \theta_{j,\,(0,0,\aaa^*)}\Big]\big[\bar p_{(0,0,\aaa^*)}+\bar p_{(0,1,\aaa^*)}+\bar p_{(1,0,\aaa^*)}\big]
+ \Big[\prod_{j>2:\,r_j=1} \theta_{j,\,(1,1,\aaa^*)}\Big]\bar p_{(1,1,\aaa^*)} \biggr\},
\end{aligned}\]
where the last equality above follows from $\theta_{j,\aaa} = \bar\theta_{j,\aaa}$ for any $j>2$.
To ensure the above equations hold, it suffices to have the following equations satisfied for any $\aaa^*\in\{0,1\}^{K-2}$
\begin{equation}\label{eq-c1}
\begin{cases}
p_{(1,1,\aaa^*)} = \bar p_{(1,1,\aaa^*)}; \\
p_{(0,0,\aaa^*)} + p_{(1,0,\aaa^*)} + p_{(0,1,\aaa^*)} = \bar p_{(0,0,\aaa^*)} +\bar p_{(1,0,\aaa^*)} +\bar p_{(0,1,\aaa^*)}.
\end{cases}
\end{equation}
\item[(b)]  For any $\rr$ such that $(r_1,r_2) = (1,0)$ and $\rr_{S_1}\neq\mz$, from \eqref{thetaeq} and the definition of the $T$-matrix,
  (\ref{eq-tpr}) can be equivalently written as
\[\begin{aligned}
&\sum_{\aaa^*}\biggr\{ \Big[\prod_{j>2:\,r_j=1} \theta_{j,\,(0,0,\aaa^*)}\Big] \big[ g_1   (p_{(0,0,\aaa^*)} + p_{(0,1,\aaa^*)}) + (1-s_1)  p_{(1,0,\aaa^*)} \big] \\
&\quad\quad\quad + \Big[\prod_{j>2:\,r_j=1} \theta_{j,\,(1,1,\aaa^*)}\Big] (1-s_1) p_{(1,1,\aaa^*)} \biggr\} \\
=~&
\sum_{\aaa^*}\biggr\{ \Big[\prod_{j>2:\,r_j=1} \theta_{j,\,(0,0,\aaa^*)}\Big] \big[ \bar g_1   (\bar p_{(0,0,\aaa^*)} + \bar p_{(0,1,\aaa^*)}) + (1-s_1)  \bar p_{(1,0,\aaa^*)} \big] \\
&\quad\quad\quad+ \Big[\prod_{j>2:\,r_j=1} \theta_{j,\,(1,1,\aaa^*)}\Big]  (1-s_1) \bar p_{(1,1,\aaa^*)} \biggr\}.
\end{aligned}\]
To ensure the above   equation holds, it suffices to have the following equations satisfied for any $\aaa^*\in\{0,1\}^{K-2}$
\begin{eqnarray}\label{eq-c2}
 \begin{cases}
   p_{(1,1,\aaa^*)}=  \bar p_{(1,1,\aaa^*)} ;\\ 
g_1   [p_{(0,0,\aaa^*)} + p_{(0,1,\aaa^*)}] + (1-s_1)  p_{(1,0,\aaa^*)}= \bar g_1  [\bar p_{(0,0,\aaa^*)} + \bar p_{(0,1,\aaa^*)}] + (1-s_1) \bar p_{(1,0,\aaa^*)}. 
 \end{cases}
\end{eqnarray}
\item[(c)]  For any $\rr$ such that $(r_1,r_2) = (0,1)$ and $\rr_{S_1}\neq\mz$, 
by symmetry to the previous case of $(r_1,r_2)=(1,0)$, when the following equations hold for any $\aaa^*\in\{0,1\}^{K-2}$, equation (\ref{eq-tpr}) is guaranteed to hold
\begin{eqnarray}\label{eq-c3}
\begin{cases}
 p_{(1,1,\aaa^*)} =   \bar p_{(1,1,\aaa^*)} ;\\ 
g_2   [p_{(0,0,\aaa^*)} + p_{(1,0,\aaa^*)}] + (1-s_2)  p_{(0,1,\aaa^*)} = \bar g_2   [\bar p_{(0,0,\aaa^*)} + \bar p_{(1,0,\aaa^*)}] + (1-s_2) \bar p_{(0,1,\aaa^*)}.
\end{cases}
\end{eqnarray}
\item[(d)]  For any $\rr$ such that $(r_1,r_2) = (1,1)$ and $\rr_{S_1}\neq\mz$, similarly to the previous cases,
equation (\ref{eq-tpr}) can be equivalently written as
\[\begin{aligned}
&\sum_{\aaa^*}\biggr\{ \Big[\prod_{j>2:\,r_j=1} \theta_{j,\,(0,0,\aaa^*)}\Big] \big[ g_1  g_2   p_{(0,0,\aaa^*)} + (1-s_1)  g_2  p_{(1,0,\aaa^*)} + g_1  (1-s_2)  p_{(0,1,\aaa^*)} \big] \\
&\quad\quad\quad+ \Big[\prod_{j>2:\,r_j=1} \theta_{j,\,(1,1,\aaa^*)}\Big]  (1-s_1)  (1-s_2)   p_{(1,1,\aaa^*)} \biggr\} \\
=&
\sum_{\aaa^*}\biggr\{ \Big[\prod_{j>2:\,r_j=1} \theta_{j,\,(0,0,\aaa^*)}\Big] \big[ \bar g_1\bar  g_2 \bar  p_{(0,0,\aaa^*)} + (1-s_1) \bar g_2 \bar p_{(1,0,\aaa^*)} +\bar  g_1 (1-s_2) \bar p_{(0,1,\aaa^*)} \big] \\
&\quad\quad\quad+ \Big[\prod_{j>2:\,r_j=1} \theta_{j,\,(1,1,\aaa^*)}\Big] (1-s_1) (1-s_2) \bar p_{(1,1,\aaa^*)}  \biggr\}.
\end{aligned}\]
 To ensure the above equation hold, it suffices to have the following equations hold for any $\aaa^*\in\{0,1\}^{K-2}$
\begin{equation}\label{eq-c4}
\begin{aligned}
\begin{cases}
&   p_{(1,1,\aaa^*)} =   \bar p_{(1,1,\aaa^*)}; \\ 
& g_1  g_2   p_{(0,0,\aaa^*)} + (1-s_1)  g_2  p_{(1,0,\aaa^*)} + g_1  (1-s_2)  p_{(0,1,\aaa^*)} \\
&\qquad=\bar g_1 \bar  g_2 \bar  p_{(0,0,\aaa^*)} + (1-s_1)  \bar g_2  \bar p_{(1,0,\aaa^*)} +\bar  g_1  (1-s_2 ) \bar p_{(0,1,\aaa^*)}.
\end{cases}
\end{aligned}
\end{equation}
\end{enumerate}

We further consider those response vectors with $\rr_{S_1}=\mz$. A similar argument gives that, to ensure (\ref{eq-tpr}) holds for any $\rr$ with $\rr_{S_1}=\mz$, it suffices  to have equations (\ref{eq-c1})--(\ref{eq-c4}) hold.
Together with the results in cases (a)--(d) discussed above, we know that   equations (\ref{eq-c1})--(\ref{eq-c4}) are a set of sufficient conditions for  (\ref{eq-tpr}) to hold  for any $\rr\in\{0,1\}^J$. Therefore, to show the necessity of Condition 2, we only need to construct      $(\bar g_1,\bar g_2,\bar\pp)\neq (  g_1,  g_2, \pp) $ satisfying (\ref{eq-c1})--(\ref{eq-c4}), which can be equivalently written as, for any $\aaa^*\in\{0,1\}^{K-2}$, $p_{(1,1,\aaa^*)} =   \bar p_{(1,1,\aaa^*)}$ and
 \begin{eqnarray}\label{eq-t1}
 \begin{cases}
& p_{(0,0,\aaa^*)} + p_{(1,0,\aaa^*)} + p_{(0,1,\aaa^*)} = \bar p_{(0,0,\aaa^*)} +\bar p_{(1,0,\aaa^*)} +\bar p_{(0,1,\aaa^*)}; \\ 
& g_1   [p_{(0,0,\aaa^*)} + p_{(0,1,\aaa^*)}] + (1-s_1)  p_{(1,0,\aaa^*)} = \bar g_1   [\bar p_{(0,0,\aaa^*)} + \bar p_{(0,1,\aaa^*)}] + (1-s_1) \bar p_{(1,0,\aaa^*)} ;\\ 
& g_2   [p_{(0,0,\aaa^*)} + p_{(1,0,\aaa^*)}] + (1-s_2)  p_{(0,1,\aaa^*)} = \bar g_2   [\bar p_{(0,0,\aaa^*)} + \bar p_{(1,0,\aaa^*)}] + (1-s_2) \bar p_{(0,1,\aaa^*)} ;\\ 
& g_1  g_2   p_{(0,0,\aaa^*)} + (1-s_1)  g_2  p_{(1,0,\aaa^*)} + g_1  (1-s_2)  p_{(0,1,\aaa^*)}  \\
&\qquad  =\bar g_1 \bar  g_2  \bar  p_{(0,0,\aaa^*)} + (1-s_1)  \bar g_2  \bar p_{(1,0,\aaa^*)} +\bar  g_1  (1-s_2)  \bar p_{(0,1,\aaa^*)}.
\end{cases}
\end{eqnarray}
 To construct      $(\bar g_1,\bar g_2,\bar\pp)\neq (  g_1,  g_2, \pp) $, we focus on the family of parameters $(\cs,\cg,\pp)$ such that for  any $\aaa^*\in\{0,1\}^{K-2}$,  
\[
\frac{p_{(0,1,\aaa^*)}}{p_{(0,0,\aaa^*)}} = u \mbox{~~and~~}
\frac{p_{(1,0,\aaa^*)}}{p_{(0,0,\aaa^*)}} = v,
\]
where $u$ and $v$ are some positive constants. Next we choose $ \bar\pp$ such that for any $\aaa^*\in\{0,1\}^{K-2}$
\[p_{(1,1,\aaa^*)} =   \bar p_{(1,1,\aaa^*)}, \quad
\bar p_{(0,0,\aaa^*)} = \bar\rho \cdot p_{(0,0,\aaa^*)},\quad
\frac{\bar p_{(0,1,\aaa^*)}}{\bar p_{(0,0,\aaa^*)}} = \bar u, \mbox{~~and~~}
\frac{\bar p_{(1,0,\aaa^*)}}{\bar p_{(0,0,\aaa^*)}} = \bar v,
\]
for some positive constants $\bar\rho$, $\bar u$ and $\bar v$ to be determined. 
In particular, we choose   $\bar\rho$ close enough to 1 
and then (\ref{eq-t1}) is equivalent  to
\begin{equation}\label{eq-t3}
\begin{aligned}
\begin{cases}
& (1+u+v) = \bar\rho  (1+\bar u+\bar v); \\ 
& g_1  (1+u) + (1-s_1) v = \bar\rho ~[~\bar g_1  (1+\bar u) + (1-s_1)  \bar v~] ;\\ 
& g_2 (1+v) + (1-s_2)  u = \bar \rho~ [~\bar g_2 (1+\bar v) + (1-s_2)  \bar u~] ;\\ 
& g_1  g_2+g_1  (1-s_2)  u+(1-s_1)  g_2  v  =\bar\rho ~ [~\bar g_1  \bar g_2+\bar g_1  (1-s_2)  \bar u+(1-s_1)  \bar g_2  \bar v~].
\end{cases}
\end{aligned}
\end{equation}
For any $g_1,g_2, s_1, s_2, u$ and $v$, the above system of equations contain 5 free parameters $\bar\rho$, $\bar u$, $\bar v$, $\bar g_1$ and $\bar g_2$, while only have 4 constraints, so there are infinitely many sets of solutions of $(\bar\rho, \bar u, \bar v, \bar g_1,  \bar g_2)$ to \eqref{eq-t3}.
This gives the non-identifiability of $(g_1, g_2,\pp)$ and hence justifies the necessity of Condition 2.
\QEDB

\paragraph{Proof of Sufficiency}
It suffices to show that if $T(\cs,\cg)\pp = T(\bar\cs,\bar\cg)\bar\pp$, then $(\cs,\cg,\pp)= ( \bar \cs,\bar\cg,\bar\pp)$.
Under Condition 1, Theorem 4 in \cite{Xu15} gives that 
$\cs =\bar\cs$  and 
$g_j = \bar g_j$ for  $j\in\{K+1,\ldots,J\}.$
It remains to show $g_j = \bar g_j$ for $j\in \{1,\ldots,K\}$.
To facilitate the proof, we introduce the following lemma, whose proof is given in the Supplementary Material.
\begin{lemma}\label{lem1}
Suppose Condition 1 is satisfied. For an item set $S$, define $\vee_{h\in S\,}\qq_h $  to be the vector of the element-wise maximum of the $\qq$-vectors in the set $S$.
For any $k\in \{1,\ldots,K\}$, if there exist two item sets, denoted by $S_k^-$ and $S_k^+$, which are not necessarily nonempty or disjoint, such that
\begin{equation}\label{eq-lem1}
\begin{aligned}
  g_h = \bar g_h \mbox{ for any } h\in S_k^-\cup S_k^+, ~and~ 
  \vee_{h\in S_k^+}\qq_h - \vee_{h\in S_k^-}\qq_h = \ee_k^\top = (\boldsymbol 0, \underbrace{1}_\text{column $k$}, \boldsymbol 0),
\end{aligned}\end{equation}
then $g_k = \bar g_k$.
\end{lemma}
Suppose the $Q$-matrix takes the form of (\ref{eq-formq}), then under Condition 2, any two different columns of the $(J-K)\times K$ sub-matrix $Q^* $ as specified in (\ref{eq-formq}) are distinct. 
Before proceeding with the proof, we first introduce the concept of the ``lexicographic order".
We denote the lexicographic order on $\{0,1\}^{J-K}$, the space of all $(J-K)$-dimensional binary vectors, by ``$\prec_{\text{lex}}$". Specifically, for any $\ba=(a_1,\ldots,a_{J-K})^\top$, $\bb=(b_1,\ldots,b_{J-K})^\top\in\{0,1\}^{J-K}$, we write $\ba \prec_{\text{lex}}\bb$ if either $a_1<b_1$; or there exists some $i\in\{2,\ldots,J-K\}$ such that $a_i<b_i$ and $a_j=b_j$ for all $j<i$.  For instance, the following four vectors $\ba_1,\ba_2,\ba_3,\ba_4$ in $\{0,1\}^2$ are sorted in an increasing lexicographic order:

\vspace{-4mm}
\singlespace
\[
\ba_1 = \begin{pmatrix} 0\\0\end{pmatrix}
\prec_{\text{lex}} 
\ba_2 = \begin{pmatrix} 0\\1\end{pmatrix}
\prec_{\text{lex}} 
\ba_3 = \begin{pmatrix} 1\\0\end{pmatrix}
\prec_{\text{lex}} 
\ba_4 = \begin{pmatrix} 1\\1\end{pmatrix}.
\]
\doublespace
It is not hard to see that if the $K$ column vectors of the submatrix $Q^*$ are mutually distinct, then there exists a unique way to sort them in an increasing lexicographic order. Thus under Condition 2, there exists a unique permutation $(k_1,k_2,\ldots,k_K)$ of $(1,2,\ldots,K)$ such that column $k_1$ has the smallest lexicographic order among the $K$ columns of $Q^*$, column $k_2$ has the second smallest lexicographic order, and so on, i.e.,
$Q^*_{\Cdot,k_1}\prec_{\text{lex}}Q^*_{\Cdot,k_2}\prec_{\text{lex}}\ldots \prec_{\text{lex}} Q^*_{\Cdot,k_K}$.
As an illustration, consider the leftmost $Q$-matrix presented in Example 1, Equation \eqref{ex3}:

\vspace{-3mm}
\singlespace
\[Q =\begin{pmatrix}
&{\cal I}_3&\\
\hdashline[2pt/2pt]
1&1&0 \\
1&0&1 \\
1&1&1 \\
1&1&1 \\
\end{pmatrix},\]
\doublespace
then the permutation is $(k_1,k_2,k_3)=(3,2,1)$, since the third column of $Q^*$ has the smallest lexicographic order while the first column has the largest.
Recall that we denote $\ba\succeq\bb$ if $a_i>b_i$ for all $i$, and denote $\ba\nsucceq\bb$ otherwise.
Then by definition, if $\ba\prec_{\text{lex}}\bb$, then $\ba\nsucceq\bb$ must hold. Therefore for any $1\leq i<j\leq K$, since $Q_{\Cdot,k_i}\prec_{\text{lex}}Q_{\Cdot,k_j}$, we must have $Q_{\Cdot,k_i}\nsucceq Q_{\Cdot,k_j}$. This fact will be useful in the following proof.

\color{black}
Equipped with the permutation $(k_1,\ldots,k_K)$,
we first prove $g_{k_1} = \bar g_{k_1}$. Define a subset of items
$$
S_{k_1}^- = \{j>K: q_{j,k_1} = 0\},
$$
which includes those items from $\{K+1,\ldots,J\}$ that do not require attribute $k_1$. Since $Q^*_{\Cdot,k_1}$ is of the smallest lexicographic order among column vectors of $Q^*$, for any $k\in\{1,\ldots,K\}\backslash \{k_1\}$, we must have
$
Q^*_{\Cdot,k} \npreceq Q^*_{\Cdot,k_1}.
$
Thus, for any $k\in\{1,\ldots,K\}\backslash \{k_1\}$ there must exist some item $j_k\in\{K+1,\ldots,J\}$ such that 
$q_{j_k,k} = 1 > 0 = q_{j_k,k_1},$
which indicates that the union of the attributes required by items in $S_{k_1}^-$ include all the attributes other than $k_1$, i.e
\[
\vee_{h\in S_{k_1}^-}\qq_h = (\mo,\underbrace{0}_\text{column $k_1$},\mo).
\]
We further define $S_{k_1}^+ = \{K+1,\ldots,J\}$. Since $S_{k_1}^-$ and $S_{k_1}^+$ satisfy conditions (\ref{eq-lem1}) in Lemma \ref{lem1} for attribute $k_1$, we have
$g_{k_1} = \bar g_{k_1}.$

Next we use the induction method to prove that for $l=2,\ldots,K$, we also have $g_{k_l}=\bar g_{k_l}$. In particular, suppose for any $1\leq m\leq l-1$, we already have $g_{k_m} = \bar g_{k_m}$. 
Note that each $k_l$ is an integer in $\{1,\ldots,K\}$ that can  be viewed as \textit{either} the index of the $k_l$th attribute \textit{or} the index of the $k_l$th item.
Define a set of items
\begin{equation}\label{eq-def}
S^{-}_{k_l} = \{j>K: q_{j,k_l} = 0\} \cup \{k_m:1\leq m\leq l-1\},
\end{equation}
where the set $\{j > K : q_{j,k_l}=0\}$ contains those items, among the last $J-K$ items, which do not require attribute $k_l$; while the set $\{k_m :1\leq m\leq l-1\}$ contains those items for which we have already established the identifiability of the guessing parameter in steps $m=1,2,\ldots,l-1$ of the induction method, i.e., $g_{k_m}=\bar g_{k_m}$ for $m=1,\ldots,l-1$.
Thus for any item $j\in S_{k_l}^-$, we have $g_{j} = \bar g_{j}$. Namely, $S_{k_l}^-$ includes the items whose guessing parameters have already been identified prior to step $l$ of the induction method.
Moreover, we claim
\begin{equation}\label{eq-vee}
\vee_{h\in S_{k_l}^-}\qq_h = (\mo,\underbrace{0}_\text{column $k_l$},\mo).
\end{equation}
This is because   for any $1\leq m\leq l-1$, the item $k_m$, whose $\qq$-vector is $\ee_{k_m}^\top$, is included in the set $S_{k_l}^-$ and hence attribute $k_m$ is required by the set $S_{k_l}^-$; 
on the other hand, for any $h\in\{l+1,\ldots, K\}$, the column vector $Q^*_{\Cdot,k_h}$ is of greater lexicographic order than $Q^*_{\Cdot,k_l}$ and hence there must exist some item in $S_{k_l}^-$ that does not require attribute $k_l$ but requires attribute $k_h$. 
We further define $S_{k_l}^+ = \{K+1,\ldots,J\}$. The chosen $S_{k_l}^-$ and $S_{k_l}^+$ satisfy the conditions (\ref{eq-lem1}) in Lemma \ref{lem1} and therefore 
$g_{k_l} = \bar g_{k_l}.$

Now  that all the slipping   and guessing parameters have been identified, $T(\cs,\cg) \pp = T(\bar\cs,\bar\cg)  \bar\pp = T(\cs,\cg)  \bar\pp$. Then the fact that $T(\cs,\cg)$ has full column rank, which is shown in  the Proof of Theorem 1 in \cite{Xu15}, implies $\pp= \bar\pp.$ 
This completes the proof.
\QEDB

\onehalfspacing
\bibliographystyle{apa}
\bibliography{bibEduc}

\begin{thebibliography}{}

\bibitem[\protect\astroncite{Casella and Berger}{2002}]{Casella}
Casella, G. and Berger, R.~L. (2002).
\newblock {\em Statistical inference}.
\newblock Duxbury Pacific Grove, CA, 2 edition.

\bibitem[\protect\astroncite{Chen et~al.}{2015}]{Chen2014}
Chen, Y., Liu, J., Xu, G., and Ying, Z. (2015).
\newblock Statistical analysis of {$Q$}-matrix based diagnostic classification
  models.
\newblock {\em Journal of the American Statistical Association}, 110:850--866.

\bibitem[\protect\astroncite{Chiu et~al.}{2009}]{Chiu}
Chiu, C.-Y., Douglas, J.~A., and Li, X. (2009).
\newblock Cluster analysis for cognitive diagnosis: theory and applications.
\newblock {\em Psychometrika}, 74:633--665.

\bibitem[\protect\astroncite{{de la Torre}}{2011}]{dela2011}
{de la Torre}, J. (2011).
\newblock The generalized {DINA} model framework.
\newblock {\em Psychometrika}, 76:179--199.

\bibitem[\protect\astroncite{DeCarlo}{2011}]{deCarlo2011}
DeCarlo, L.~T. (2011).
\newblock On the analysis of fraction subtraction data: the {DINA} model,
  classification, class sizes, and the {Q}-matrix.
\newblock {\em Applied Psychological Measurement}, 35:8--26.

\bibitem[\protect\astroncite{DiBello et~al.}{1995}]{DiBello}
DiBello, L.~V., Stout, W.~F., and Roussos, L.~A. (1995).
\newblock Unified cognitive psychometric diagnostic assessment likelihood-based
  classification techniques.
\newblock In Nichols, P.~D., Chipman, S.~F., and Brennan, R.~L., editors, {\em
  Cognitively diagnostic assessment}, pages 361--390. Erlbaum Associates,
  Hillsdale, NJ.

\bibitem[\protect\astroncite{Fang et~al.}{2017}]{GLY}
Fang, G., Liu, J., and Ying, Z. (2017).
\newblock On the identifiability diagnostic classification models.
\newblock arXiv Preprint.

\bibitem[\protect\astroncite{Gabrielsen}{1978}]{gabrielsen1978consistency}
Gabrielsen, A. (1978).
\newblock Consistency and identifiability.
\newblock {\em Journal of Econometrics}, 8(2):261--263.

\bibitem[\protect\astroncite{Goodman}{1974}]{GOODMAN74}
Goodman, L.~A. (1974).
\newblock Exploratory latent structure analysis using both identifiable and
  unidentifiable models.
\newblock {\em Biometrika}, 61:215--231.

\bibitem[\protect\astroncite{Henson et~al.}{2009}]{HensonTemplin09}
Henson, R.~A., Templin, J.~L., and Willse, J.~T. (2009).
\newblock Defining a family of cognitive diagnosis models using log-linear
  models with latent variables.
\newblock {\em Psychometrika}, 74:191--210.

\bibitem[\protect\astroncite{Junker and Sijtsma}{2001}]{Junker}
Junker, B.~W. and Sijtsma, K. (2001).
\newblock Cognitive assessment models with few assumptions, and connections
  with nonparametric item response theory.
\newblock {\em Applied Psychological Measurement}, 25:258--272.

\bibitem[\protect\astroncite{Koopmans and Reiers\o{}l}{1950}]{Koopmans50}
Koopmans, T.~C. and Reiers\o{}l, O. (1950).
\newblock The identification of structural characteristics.
\newblock {\em Ann. Math. Statist.}, 21:165--181.

\bibitem[\protect\astroncite{Liu et~al.}{2013}]{JLGXZY2011}
Liu, J., Xu, G., and Ying, Z. (2013).
\newblock Theory of self-learning {Q}-matrix.
\newblock {\em Bernoulli}, 19(5A):1790--1817.

\bibitem[\protect\astroncite{Maris and Bechger}{2009}]{MarisBechger}
Maris, G. and Bechger, T.~M. (2009).
\newblock Equivalent diagnostic classification models.
\newblock {\em Measurement}, 7:41--46.

\bibitem[\protect\astroncite{McHugh}{1956}]{McHugh}
McHugh, R.~B. (1956).
\newblock Efficient estimation and local identification in latent class
  analysis.
\newblock {\em Psychometrika}, 21:331--347.

\bibitem[\protect\astroncite{Rothenberg}{1971}]{rothenberg1971identification}
Rothenberg, T.~J. (1971).
\newblock Identification in parametric models.
\newblock {\em Econometrica}, 39:577--591.

\bibitem[\protect\astroncite{Tatsuoka}{2009}]{TatsuokaC09}
Tatsuoka, C. (2009).
\newblock Diagnostic models as partially ordered sets.
\newblock {\em Measurement}, 7:49--53.

\bibitem[\protect\astroncite{{Tatsuoka}}{1983}]{Tatsuoka1983}
{Tatsuoka}, K.~K. (1983).
\newblock Rule space: an approach for dealing with misconceptions based on item
  response theory.
\newblock {\em Journal of Educational Measurement}, 20:345--354.

\bibitem[\protect\astroncite{Templin and Henson}{2006}]{Templin}
Templin, J.~L. and Henson, R.~A. (2006).
\newblock Measurement of psychological disorders using cognitive diagnosis
  models.
\newblock {\em Psychological Methods}, 11:287--305.

\bibitem[\protect\astroncite{{von Davier}}{2005}]{von}
{von Davier}, M. (2005).
\newblock A general diagnostic model applied to language testing data.
\newblock Research report, Educational Testing Service, Princeton, NJ.

\bibitem[\protect\astroncite{von Davier}{2014}]{davier2014dina}
von Davier, M. (2014).
\newblock The {DINA} model as a constrained general diagnostic model: {T}wo
  variants of a model equivalency.
\newblock {\em British Journal of Mathematical and Statistical Psychology},
  67(1):49--71.

\bibitem[\protect\astroncite{Wang and Douglas}{2015}]{wang2015consistency}
Wang, S. and Douglas, J. (2015).
\newblock Consistency of nonparametric classification in cognitive diagnosis.
\newblock {\em Psychometrika}, 80(1):85--100.

\bibitem[\protect\astroncite{Xu}{2017}]{Xu2016}
Xu, G. (2017).
\newblock Identifiability of restricted latent class models with binary
  responses.
\newblock {\em The Annals of Statistics}, 45:675--707.

\bibitem[\protect\astroncite{Xu and Shang}{2017}]{Xu2017}
Xu, G. and Shang, Z. (2017).
\newblock Identifying latent structures in restricted latent class models.
\newblock {\em Journal of the American Statistical Association}, (accepted).

\bibitem[\protect\astroncite{Xu and Zhang}{2016}]{Xu15}
Xu, G. and Zhang, S. (2016).
\newblock Identifiability of diagnostic classification models.
\newblock {\em Psychometrika}, 81:625--649.

\end{thebibliography}

\newpage
\doublespacing
\noindent
{\LARGE \textbf{Supplementary Material}}

\appendix
 \section*{A1: Derivation of Equation \eqref{a^2} in Example 2}
In Example 2, we claimed that, given the $Q$-matrix in the following form where there are $J_0$ items with $\bq$-vectors being $(0,0)$ and $J-2-J_0$ items with $\bq$-vectors being $(1,1)$,

\vspace{-4mm}
\singlespacing
\begin{equation*}
Q=\begin{pmatrix}
 {\cal I}_2  \\
0\quad 0\\
\vdots\quad \vdots\\
0\quad 0\\
1\quad 1\\
\vdots\quad \vdots\\
1\quad 1\\
\end{pmatrix}_{J\times 2},
\end{equation*}
\doublespacing

\noindent
to construct $(\bar\cs,\bar\cg,\bar\pp)\neq (\cs,\cg,\pp)$ satisfying Equation \eqref{eq-orig} where $\bar\cs=\cs$, $\bar g_j = g_j$ for all $j=3,\ldots,J$, and $\bar p_{(1,1)} = p_{(1,1)}$, it suffices to ensure the Equations \eqref{a^2} hold.
Now we prove this argument. Following the proof of the necessity of  Conditions C1 and C2 in the Appendix, we can obtain the following equations in   \eqref{eq-st1} from Equations \eqref{eq-t1} in the main text by   replacing $(\alpha_1,\alpha_2,\aaa^*)$ in  \eqref{eq-t1} with $(\alpha_1,\alpha_2)$ here, since in this case there are only two attributes. And similarly we have the conclusion that Equation \eqref{eq-orig} holds as long as   Equations \eqref{eq-st1} hold,

\begin{equation}\label{eq-st1}
\begin{cases}
 p_{(0,0)} + p_{(1,0)} + p_{(0,1)} = \bar p_{(0,0)} +\bar p_{(1,0)} +\bar p_{(0,1)}; \\ 
 g_1   [p_{(0,0)} + p_{(0,1)}] + (1-s_1)  p_{(1,0)} = \bar g_1   [\bar p_{(0,0)} + \bar p_{(0,1)}] + (1-s_1) \bar p_{(1,0)} ;\\ 
g_2   [p_{(0,0)} + p_{(1,0)}] + (1-s_2)  p_{(0,1)} = \bar g_2   [\bar p_{(0,0)} + \bar p_{(1,0)}] + (1-s_2) \bar p_{(0,1)} ;\\ 
 g_1  g_2   p_{(0,0)} + (1-s_1)  g_2  p_{(1,0)} + g_1  (1-s_2)  p_{(0,1)}  \\
\qquad =\bar g_1 \bar  g_2  \bar  p_{(0,0)} + (1-s_1)  \bar g_2  \bar p_{(1,0)} +\bar  g_1  (1-s_2)  \bar p_{(0,1)}.
\end{cases}
\end{equation}

\doublespacing

\noindent
Adding $p_{(1,1)}$ to both hand sides of the first equation in \eqref{eq-st1}, adding $(1-s_1)p_{(1,1)}$ to the second equation, adding $(1-s_2)p_{(1,1)}$ to the third equation and adding $(1-s_1)(1-s_2)p_{(1,1)}$ to the last equation, we exactly obtain \eqref{a^2} in Example 2.

\color{black}
\section*{A2: Proof of Corollary \ref{cor1}} 
	 When the identifiability conditions are satisfied, the maximum likelihood estimators of $\hat\cs,\hat\cg$, and $\hat\pp$ are consistent as the sample size $N\to\infty$. Specifically,
  we introduce  a $2^J$-dimensional empirical response vector 
\begin{eqnarray*}
	\boldsymbol\gamma
	&= & \biggr\{
1,
{N}^{-1}\sum_{i=1}^NI(\rr_i \succeq \ee_1),
\cdots,
{N}^{-1}\sum_{i=1}^NI(\rr_i \succeq \ee_J),\\
&&\quad 
{N}^{-1}\sum_{i=1}^NI(\rr_i \succeq \ee_1+\ee_2),
\cdots,
{N}^{-1}\sum_{i=1}^NI(\rr_i \succeq \mathbf 1)\biggr\}^\top,
\end{eqnarray*}
where elements of $\boldsymbol\gamma$ are indexed by  response vectors  arranged in the same order as the rows of the $T$-matrix.  
 From the definition of the $T$-matrix and the law of large numbers, we know
 $\boldsymbol\gamma\to
T(\cs,\cg) \pp  
 $ almost surely as $N\to\infty$. On the other hand, the maximum likelihood estimators $\hat\cs,\hat\cg$, and $\hat\pp$ satisfy 
$\|\boldsymbol\gamma - 
T(\hat\cs,\hat\cg) \hat\pp\| \to 0 , $ where $\|\cdot\|$ is the $L_2$ norm.
Therefore,  
\begin{equation*}
\|T( \cs, \cg) \pp - 
T(\hat\cs,\hat\cg) \hat\pp\| \to 0  
\end{equation*}  almost surely.  Then from the proof of Theorem 1, we can obtain the consistency result  that $(\hat\cs,\hat\cg,\hat\pp)\to (\cs,\cg,\pp)$ almost surely as $N\to\infty$. 
 \QEDB
 
 \section*{A3: Proof of Proposition \ref{prop-q0}}
Consider a $Q$-matrix of size $J\times K$ in the form
\begin{equation}\label{eq-qq}
Q=\begin{pmatrix}
Q'\\
\mathbf0\\
\end{pmatrix},\end{equation}
where $Q'$ is of size $J'\times K$ and contains those nonzero $\bq$-vectors of $Q$.
Recall from the model setup in Section 2 of the main text, for any item   $j\in\{J'+1,\ldots,J\}$ which has   $\bq_j=\mathbf 0$, the guessing parameter is not needed by the DINA model and for notational convenience, we set $g_j \equiv\bar g_j \equiv 0$, so the slipping parameter $s_j$ is the only unknown item parameter associated with such $j$. 
Taking the response pattern $\rr=\ee_j$ for any item $j\in\{J'+1,\ldots,J\}$ in Equation \eqref{eq-tpr} gives
\[T_{\ee_j,\Cdot}( \cs,\cg) \pp = (1-s_j) \sum_{\aaa\in\{0,1\}^K} p_{\aaa}
=(1-\bar s_j) \sum_{\aaa\in\{0,1\}^K} \bar p_{\aaa}
= T_{\ee_j,\Cdot}( \bar\cs,\bar\cg) \bar\pp,\]
then since $\sum_{\aaa\in\{0,1\}^K}  p_{\aaa}=\sum_{\aaa\in\{0,1\}^K} \bar p_{\aaa}=1$,  we have $s_j=\bar s_j$ for any $j\in\{J'+1,\ldots,J\}$. 

Now denote $\cs'=(s_1,\ldots,s_{J'})$, $\cg'=(g_1,\ldots,g_{J'})$ and similarly denote $\bar\cs'$, $\bar\cg'$. Denote the $2^{J'}\times 2^K$ $T$-matrix associated with matrix $Q'$ by $T'(\cs',\cg')$.
For any response pattern $\rr=(r_1,\ldots,r_{J'},r_{J'+1},\ldots,r_J)\in\{0,1\}^J$, denote 
$ \rr' =(r_1,\ldots,r_{J'})$
and $(\rr',\mathbf0)=(r_1,\ldots,r_{J'},0,\ldots,0)$ of length $J$;
 then we have 
\[\begin{aligned}
T_{\rr,\Cdot}( \cs,\cg)\pp &=
\Big\{T_{(\rr',\mathbf0),\Cdot}( \cs,\cg)\pp\Big\} \prod_{j>J'}(1-s_j)^{r_j}
=\Big\{T'_{\rr',\Cdot}( \cs',\cg')\pp\Big\} \prod_{j>J'}(1-s_j)^{r_j},\\
T_{\rr,\Cdot}(\bar \cs,\bar\cg)\bar\pp &=
\Big\{T_{(\rr',\mathbf0),\Cdot}( \bar\cs,\bar\cg)\bar\pp\Big\} \prod_{j>J'}(1-s_j)^{r_j}
=\Big\{T'_{\rr',\Cdot}( \bar\cs',\bar\cg')\pp\Big\} \prod_{j>J'}(1-s_j)^{r_j}.
\end{aligned}\]
Using the above equalities, by Proposition \ref{equiv}, we have the following equivalent arguments,
\[\begin{aligned}
&\quad (\cs,\cg,\pp)~\mbox{associated with}~ Q~\mbox{are identifiable,}\\
\Longleftrightarrow &\quad\forall (\bar\cs,\bar\cg,\bar\pp)\neq (\cs,\cg,\pp),~ 
\exists\rr\in\{0,1\}^{J}~\mbox{such that}~T_{\rr,\Cdot}( \cs,\cg) \pp\neq T_{\rr,\Cdot}(  \bar\cs,\bar\cg)\bar\pp,\\
\Longleftrightarrow &\quad\forall (\bar\cs,\bar\cg,\bar\pp)\neq (\cs,\cg,\pp), 
~\exists\rr'\in\{0,1\}^{J'}~\mbox{such that}~
T'_{\rr',\Cdot}( \cs',\cg') \pp\neq T'_{\rr',\Cdot}(  \bar\cs',\bar\cg')\bar\pp,\\
\Longleftrightarrow&\quad (\cs',\cg',\pp)~\mbox{associated with}~ Q'~\mbox{are identifiable.}
\end{aligned}\] 
Therefore we have shown  identifiability of DINA associated with $Q$ in the form of \eqref{eq-qq} is equivalent to that of DINA associated with submatrix $Q'$ in \eqref{eq-qq} and the proof of the proposition is complete.


\color{black}
\section*{A4: Proof of Lemma \ref{lem1}}
To facilitate the proof of the lemma, we introduce the following proposition, which is from \ Proposition 3 in \cite{Xu2016}.
We first generalize the definition of the $T$-matrix. For any $\ca=(x_1,\ldots,x_J)^\top\in\mathbb R^J$ and $\cb=(y_1,\ldots,y_J)^\top\in\mathbb  R^J$, we still define the $T$-matrix $T(\ca, \cb)$ to be a $2^J\times 2^K$ matrix, where the entries are indexed by row index $\rr\in \{0,1\}^J$ and column index $\aaa$.
For any row indexed by $\ee_j$ with $j=1,\ldots,J$, we let $t_{\ee_j,\aaa}(\ca,\cb) =(1-x_j)^{\xi_{j,\aaa} }y_j^{1-\xi_{j,\aaa}}$; for any $\rr\neq\mz$, let the $\rr$th row vector of $T(\ca, \cb)$  be
$T_{\rr,\Cdot}( \ca,\cb) = \bigodot_{j: r_j=1} T_{\ee_j,\Cdot}( \ca,\cb)$. 

\begin{proposition}\label{prop-ltrans}
If $T(\cs ,\cg )\pp = T(\bar\cs ,\bar\cg )\bar\pp$, then  for any $\ttt\in\mathbb  R^J$,
$
T(\cs+\ttt,\cg-\ttt)\pp = T(\bar\cs+\ttt,\bar\cg-\ttt)\bar\pp.
$ 
\end{proposition}

Let $G$ be the set of items whose guessing parameters have been identified in the sense that $g_j=\bar g_j,$ for any $j\in G$. Let $G^c:=\{1,\ldots,J\}\backslash G$ be the complement of $G$.
Note that $\{K+1,\ldots,J\}\cup S_k^-\cup S_k^+\subseteq G.$
Define \begin{equation}\label{eq-theta}
\ttt = \sum_{j\in G^c}(1-s_j)\ee_j + \sum_{j\in G}g_j\ee_j.
\end{equation}
 Denote  $T  := T(\cs=\mz,\cg=\mz)$  and denote the $(\rr,\aaa)$-entry of $T$ by $t_{\rr,\aaa}$, then by definition, 
\begin{equation}\label{eq-tra}
t_{\rr,\aaa}=\prod_{j:\,r_j=1}1^{I(\aaa\succeq \qq_j)}0^{1-I(\aaa\succeq \qq_j)}=I(\aaa\succeq \qq_j~\forall j~\text{s.t.}~r_j=1),
\end{equation}
where $I(\cdot)$ denotes the   indicator function. Proposition \ref{prop-ltrans} implies that $T_{\rr,\Cdot}(\cs+\ttt,\cg-\ttt)=T_{\rr,\Cdot}(\cs+\ttt,\bar\cg-\ttt) \bar\pp$ for $\ttt$ defined in \eqref{eq-theta}. We use $\theta_{j,\aaa}$ to denote the positive response probability of attribute profile $\aaa$ to item $j$, i.e.,  $\theta_{j,\aaa}=1-s_j$ for $\aaa$ such that $\aaa\succeq \qq_j$, and $\theta_{j,\aaa}=g_j$ for $\aaa$ such that $\aaa\nsucceq \qq_j$.
For any response pattern $\rr$ such that $r_j = 0$ for all $j\in G^c$,
\begin{equation}\label{eq-denom}
\begin{aligned}
T_{\rr,\Cdot}(\cs+\ttt,\cg-\ttt)  \pp 
&= \sum_{\aaa\in\{0,1\}^K} p_{\aaa} \prod_{j\in G} [\theta_{j,\aaa} - g_j]^{r_j} \prod_{j\in G^c} [\theta_{j,\aaa} - (1-s_j)]^{r_j} \\
&= \sum_{\aaa\in\{0,1\}^K} p_{\aaa} \prod_{j\in G} (\theta_{j,\aaa} - g_j)^{r_j},\\
\end{aligned}
\end{equation}
where in the above summation over $\aaa\in\{0,1\}^{K}$, one can see that the product term $\prod_{j\in G} (\theta_{j,\aaa} - g_j)^{r_j}$ is nonzero  only for those $\aaa$ such that $\theta_{j,\aaa}=1-s_j>g_j$  for all $j$ where $r_j=1$; and when the product term is nonzero, it equals $\prod_{j\in G} (1-s_j - g_j)^{r_j}$. 
Further examining those $\aaa$ that make the product term nonzero in \eqref{eq-denom}, one can find it is exactly those $\aaa$ such that $t_{\rr,\aaa}=1$ according to \eqref{eq-tra}. Noting that $t_{\rr,\aaa}$ can either be 1 or 0, \eqref{eq-denom} can be further written as
\begin{equation}\label{eq-denom2}
\begin{aligned}
T_{\rr,\Cdot}(\cs+\ttt,\cg-\ttt)  \pp =&\sum_{\aaa:\,t_{\rr,\aaa}=1} p_{\aaa} \prod_{j\in G} (1-s_j - g_j)^{r_j}\\
=& \sum_{\aaa\in\{0,1\}^K} t_{\rr,\aaa}  p_{\aaa}\prod_{j\in G} (1-s_j - g_j)^{r_j}. \\
\end{aligned}
\end{equation}
Following the same argument, we also have $$T_{\rr,\Cdot}(\cs+\ttt,\bar\cg-\ttt) \bar\pp=\sum_{\aaa\in\{0,1\}^K} t_{\rr,\aaa}\bar  p_{\aaa}\prod_{j\in G} (1-s_j - g_j)^{r_j},$$
then Proposition \ref{prop-ltrans} implies
\begin{equation}\label{eq-tp}
\sum_{\aaa\in\{0,1\}^K}t_{\rr,\aaa} p_{\aaa} 
=
\sum_{\aaa\in\{0,1\}^K}t_{\rr,\aaa} \bar p_{\aaa},\mbox{ for any } \rr\text{ such that } r_{j}=0\text{ for all } j\in G^c.
\end{equation}
\color{black}
We then define a response vector $\rr^*=(r^*_1,\ldots,r^*_J)^\top$  to be
$\rr^* = \sum_{j\in G} (1-q_{j,k})\ee_j,$
that is, $\rr^*$ has correct responses to and only to those items among the set $G$ that do not require the $k$th  attribute. Let $S_{\rr^*}$ denote the set of items that $\rr^*$ has correct responses to, i.e., $S_{\rr^*} = \{j: r^*_j=1\}$.
Since $S_k^- \subseteq G$ and $q_{j,k}=0$ for any $j\in S_k^-$, we know $S_{\rr^*}$ is nonempty.  
Now consider the row vector in the transformed $T$-matrix $T(\cs+\ttt,\cg-\ttt)$ corresponding to response vector $\rr^*+\ee_k$, then we have that
$T_{\rr^*+\ee_k,\aaa}(\cs+\ttt,\cg-\ttt) \neq 0$ if and only if  
$$\aaa\succeq \qq_j \mbox{ for any item } j\in S_{\rr^*}, \mbox{ and }\alpha_k=0.$$ In other words, $T_{\rr^*+\ee_k,\aaa}(\cs+\ttt,\cg-\ttt) \neq 0$ if and only if $\aaa$ satisfies $t_{\rr^*,\aaa}=1$ and $ t_{\rr^*+\ee_k,\aaa}=0$.
This implies that 
\begin{equation}\label{eq-num1}
\begin{aligned}
&T_{\rr^*+\ee_k,\Cdot}(\cs+\ttt,\cg-\ttt)\pp \\
&=
(g_k+s_k-1)\prod_{j\in S_{\rr^*}} (1-s_j-g_j) \sum_{\aaa\in\{0,1\}^K} 
   (t_{\rr^*,\aaa} - t_{\rr^*+\ee_k,\aaa}) p_{\aaa}
\end{aligned}\end{equation}
and
\begin{equation}\label{eq-num2}
\begin{aligned}
&T_{\rr^*+\ee_k,\Cdot}(Q,\cs+\ttt,\bar\cg-\ttt)\cdot \bar\pp \\
&=
(\bar g_k+s_k-1)\prod_{j\in S_{\rr^*}} (1-s_j-g_j) \sum_{\aaa\in\{0,1\}^K} 
   (t_{\rr^*,\aaa} - t_{\rr^*+\ee_k,\aaa}) \bar p_{\aaa}.
\end{aligned}\end{equation}
Note that  \eqref{eq-num1} = \eqref{eq-num2} by Proposition 2.

We next show that the summation terms in \eqref{eq-num1} and \eqref{eq-num2} satisfy  
\begin{equation}\label{eq-num3}
	\sum_{\aaa\in\{0,1\}^K} 
   (t_{\rr^*,\aaa} - t_{\rr^*+\ee_k,\aaa}) p_{\aaa} =\sum_{\aaa\in\{0,1\}^K} 
   (t_{\rr^*,\aaa} - t_{\rr^*+\ee_k,\aaa}) \bar p_{\aaa} \neq 0.
\end{equation}
Note $\rr^*$ satisfies the condition in \eqref{eq-tp} that $r_{j}^*=0$  for all $ j\in G^c.$ Therefore,
\begin{equation}\label{eq-tp2}\sum_{\aaa\in\{0,1\}^K}t_{\rr^*,\aaa} p_{\aaa} 
=
\sum_{\aaa\in\{0,1\}^K}t_{\rr^*,\aaa} \bar p_{\aaa}.
\end{equation}
We further consider the  response vector $\rr^*+\ee_k$. Under the conditions of   Lemma 1, there exists some item $h\in G$  such that 
 $$q_{h,k}=1 \mbox{ and } \{l: q_{h,l}=1,~ l\neq k\}\subseteq  \bigcup_{j\in S_{\rr^*}} \{l: q_{j,l}=1 \}.$$
That is, the item $h$ requires  the $k$th attribute  and $h$'s any other required  attribute is also required by some item in the set  $S_{\rr^*}$.    Therefore we have 
$T_{\rr^*+\ee_k,\Cdot}=T_{\rr^*\vee\rr^{\#},\Cdot }$, where   $\rr^{\#} := \sum_{h\in S_j^+\backslash S_j^-}\ee_h$; in addition, since the response vector $\rr^*\vee\rr^{\#}$ satisfies the condition in (\ref{eq-tp}) that its $j$th element $(\rr^*\vee\rr^{\#})_j=0$ for any $j\in G^c$, we have
\begin{equation}\label{eq-tp3}\begin{aligned}
  \sum_{\aaa\in\{0,1\}^K} 
   t_{\rr^*+\ee_k,\aaa}\cdot p_{\aaa}
&= \sum_{\aaa\in\{0,1\}^K} t_{\rr^*\vee\rr^{\#},\aaa}\cdot p_{\aaa}\\
&= \sum_{\aaa\in\{0,1\}^K} t_{\rr^*\vee\rr^{\#},\aaa}\cdot\bar p_{\aaa}
= \sum_{\aaa\in\{0,1\}^K} 
   t_{\rr^*+\ee_k,\aaa}\cdot \bar p_{\aaa}.
\end{aligned}\end{equation}
The first equation in \eqref{eq-num3} then follows from \eqref{eq-tp2} and \eqref{eq-tp3}. The inequality in  \eqref{eq-num3} also holds since $t_{\rr^*,\aaa}\geq t_{\rr^*+\ee_k,\aaa}$ for any $\aaa$ and $t_{\rr^*,\aaa}> t_{\rr^*+\ee_k,\aaa}$ for those $\aaa$ with $\alpha_k=0$ and $\aaa\succeq \qq_j$ for any item $j\in S_{\rr^*}$.

With the results in \eqref{eq-num3}, we have 
$g_k 
= \bar g_k$ from the equality of  \eqref{eq-num1} and \eqref{eq-num2}.
This completes the proof.
\QEDB

\end{document}